\documentclass[11pt]{article}
\usepackage{}
\usepackage{amsfonts}
\usepackage{amsmath,amssymb}
\usepackage{graphicx}
\usepackage{caption2}
\usepackage{epsfig}
\usepackage{subfigure}

\def\rm{\mathrm}

\def\dref#1{(\ref{#1})}

\begin{document}

\begin{center}
{\Large\bf{Distributed Adaptive Attitude Synchronization of Multiple
Spacecraft}} \footnote[1] {\small Zhongkui Li is with the School of Automation,
Beijing Institute
of Technology, Beijing 100081, P. R. China (E-mail:
zhongkli@gmail.com). Zhisheng Duan is with with State Key Lab for Turbulence and Complex Systems,
Department of Mechanics and Aerospace Engineering, College of Engineering, Peking University,
Beijing 100871, P. R. China (E-mail:
duanzs@pku.edu.cn)}
\end{center}
\vskip 0.2cm
\centerline{ Zhongkui Li and Zhisheng Duan}
\vskip 0.7cm

{\noindent \small {\bf  Abstract}: This paper addresses the distributed
attitude synchronization problem of multiple spacecraft with unknown inertia matrices.
Two distributed adaptive controllers are proposed for the cases with and without a virtual
leader to which a time-varying reference attitude is assigned.
The first controller achieves attitude synchronization for a group of spacecraft with a
leaderless communication topology having a directed spanning tree.
The second controller guarantees that all spacecraft track the reference
attitude if the virtual leader has a directed path
to all other spacecraft. Simulation examples are presented to illustrate the
effectiveness of the results.

\vskip 0.2cm

}.

\section{Introduction}

In recent years, consensus and cooperative control of multi-vehicle systems have attracted
compelling attention from various scientific communities. A large body of theoretical advances
has been reported, see
\cite{li2, li2009h, olfati-saber2004consensus,ren05,ren07, lin2008distributed} and references therein.
In the aforementioned works, the agent dynamics are restricted to be a single, double integrators
or linear systems. The results proposed in these papers become quite limited when
dealing with the attitude synchronization problem of multiple spacecraft, which
 is more challenging than the consensus of vehicles with integrator dynamics, due
to the nonlinearity of the attitude dynamics.

Attitude control of a single rigid body has been extensively
studied, e.g., in \cite{wen,tsi98,ake01,wong}. A leader-follower
strategy is proposed in \cite{wang} for attitude synchronization of
multiple spacecraft. Decentralized control laws using the
behavioral approach are presented for attitude synchronization in
\cite{ren071,law02}, where the communication topology among
spacecraft is assumed to be a bidirectional ring. Adaptive
consensus protocols are proposed in \cite{cheng} for multiple
manipulators with uncertain dynamics. Cooperative attitude control
of multiple rigid bodies with a leader-follower communication
topology is considered in \cite{dim09}. In \cite{chung}, contraction
analysis theory is used to derive attitude synchronization
strategies with global exponential convergence for a group of
spacecraft with a bidirectional communication topology. Distributed
control laws without velocity measurements are studied for attitude
synchronization of multiple spacecraft in \cite{abd09} by use of
quaternion representation while in \cite{ren10} by use of Modified
Rodriguez Parameters (MRPs) for attitude representation. The results
in \cite{abd09,ren10} are applicable to general undirected
communication topologies. The distributed attitude tracking problem
is addressed in \cite{ren072} for spacecraft whose information
exchange graph can be simplified to a graph with only one node.

Motivated by \cite{ren10,chung,cheng}, this paper concerns the
distributed adaptive attitude synchronization problem of a group of
spacecraft with unknown inertia matrices. The attitude dynamics are
represented here by MRPs. Two distributed adaptive controllers are
proposed for the cases with and without a virtual leader to which a
time-varying reference attitude is assigned. The first controller
achieves attitude synchronization for a group of spacecraft with a
leaderless communication topology having a directed spanning tree.
The second controller guarantees that all spacecraft track the
reference attitude which is available to only a subset of the
spacecraft, if the virtual leader has a directed path to all other
spacecraft. Differing from the results given in \cite{chung} which
applies only to a bidirectional ring communication topology, and
those in \cite{cheng} which requires the communication graph to be
undirected, the communication topology among the spacecraft is
relaxed to a general directed graph in this paper. The results
obtained here generalize Theorem 4.1 in [17] to the case where the
attitude dynamics are uncertain and the communication topology is
either leaderless or leader-follower. It should be mentioned that
all the results in this paper are applicable to robotic manipulators
with dynamics represented by the Euler-Lagrange equation.

The rest of this paper is organized as follows. The attitude
dynamics and some useful results of the graph theory are introduced
in Section 2. The distributed adaptive attitude synchronization
problem for the cases without and with a reference attitude are
studied in Sections 3 and 4, respectively. Simulation
examples are presented to illustrate the theoretical results in
Section 5. Section 6 concludes the paper.

The following notation will be used throughout the paper.
$\mathbf{R}^{n\times n}$ denotes the set of all $n\times n$ real
matrices. $I_N$ represents the identity matrix of dimension $N$.
$\mathbf{1}\in \mathbf{R}^p$ denotes the vector with all entries
equal to one. Matrices, if not explicitly stated, are assumed to
have compatible dimensions. $\|M\|$ represents the induced 2-norm of
matrix $M\in\mathbf{R}^{n\times m}$. For a vector
$x=[x_1,x_2,x_3]^T\in\mathbf{R}^3$,
the cross-product operator is denoted by $S(x)=\begin{bmatrix}0 & -x_3 & x_2\\
x_3 & 0 & -x_1\\ -x_2 & x_1 & 0\end{bmatrix}$.
$\rm{diag}(A_1,\cdots,A_n)$ represents a block-diagonal matrix with
matrices $A_i,i=1,\cdots,n,$ on its diagonal. $A\otimes B$ denotes
the Kronecker product of matrices $A$ and $B$. 

\section{Preliminaries and Problem Formulation}

This paper considers the attitude synchronization problem
of a network of $N$ spacecraft. Modified Rodriguez Parameters (MRPs)
are used here
to represent the attitude of the spacecraft with respect to the
inertial frame. The MRP vector $\sigma_i\in\mathbf{R}^3$ for the
$i$-th spacecraft is defined by
$\sigma_i=\hat{e}_i\tan(\frac{\phi_i}{4})$, where $\hat{e}_i$ is the
Euler axis and $\phi_i$ is the Euler angle \cite{shu93}. The attitude dynamics of the
$i$-th
spacecraft are given by \cite{tsi98}
\begin{equation}\label{rbd}
\begin{aligned}
J_i\dot{\omega}_i &=-S(\omega_i)J_i\omega_i+u_i,\\
\dot{\sigma}_i &=G(\sigma_i)\omega_i,\quad i=1,2,\cdots,N,
\end{aligned}
\end{equation}
where $\omega_i\in\mathbf{R}^3$ denotes the angular velocity
in the body-fixed frame, $J_i\in\mathbf{R}^{3\times 3}$ is the inertia matrix,
$u_i\in\mathbf{R}^3$ is the control torque,
and $$G(\sigma_i)=\frac{1}{2}\left(\frac{1-\sigma_i^T\sigma_i}{2}I_3-S(\sigma_i)
+\sigma_i\sigma_i^T\right).$$

Following \cite{slotine,wong}, \dref{rbd} can be rewritten as
\begin{equation}\label{rb}
H^*_i(\sigma_i)\ddot{\sigma}_i+C^*_i(\sigma_i,\dot{\sigma}_i)\dot{\sigma}_i=
G^{-T}(\sigma_i)u_i,\quad i=1,2,\cdots,N,
\end{equation}
where
$$\begin{aligned}
H^*_i(\sigma_i)&\triangleq G^{-T}(\sigma_i)J_iG^{-1}(\sigma_i),\\
C^*_i(\sigma_i,\dot{\sigma}_i)&\triangleq
G^{-T}(\sigma_i)J_iG^{-1}(\sigma_i)\dot{G}(\sigma_i)G^{-1}(\sigma_i)
-G^{-T}(\sigma_i)S(J_iG^{-1}(\sigma_i)\dot{\sigma}_i)G^{-1}(\sigma_i).
\end{aligned}$$

It is assumed that the inertia matrices $J_i$, $i=1,2,\cdots,N$, are
unknown constant positive-definite matrices. Under this assumption,
the attitude dynamics \dref{rb} has the following properties:

{\bf \small Property 1}. Matrix $H^*_i(\sigma_i)$ is symmetric and positive definite.

{\bf \small Property 2}. Matrix
$\dot{H}^*_i(\sigma_i)-2C^*_i(\sigma_i,\dot{\sigma}_i)$ is skew
symmetric, i.e.,
$$x^T(\dot{H}^*_i(\sigma_i)-2C^*_i(\sigma_i,\dot{\sigma}_i))x=0,\quad
\forall\,x\in\mathbf{R}^3.$$

{\bf \small Property 3}. The attitude dynamics \dref{rb} satisfies
the following linear parameterization condition:
$$H^*_i(\sigma_i)\ddot{y}+C^*_i(\sigma_i,\dot{\sigma}_i)\dot{y}=
Y(\sigma_i,\dot{\sigma}_i,\dot{y},\ddot{y})\,\theta_i,\quad
\forall\,y\in\mathbf{R}^3,$$ where $Y\in\mathbf{R}^{3\times 6}$ is called the
regression matrix and
\begin{equation}\label{lpm}
\theta_i=\begin{bmatrix}J^i_{11} &
J^i_{12} & J^i_{13} & J^i_{22} & J^i_{23} &
J^i_{33}\end{bmatrix}^T
\end{equation}
is the unknown parameter vector
with $J^i_{jk}$ being the $(j,k)$-th entry of the inertia matrix
$J_i$ in \dref{rbd}.

The communication topology among spacecraft is represented by a directed graph $\mathcal {G}$
consisting of a node set $\mathcal {V}=\{1,2,\cdots,N\}$ and an edge set
$\mathcal {E}\subset\mathcal {V}\times\mathcal {V}$. The node $i$ denotes the $i$-th spacecraft.
An edge $(i,j)$ means that spacecraft $j$ can obtain the attitude information of spacecraft $i$,
but not conversely. For an edge $(i,j)$ in the directed graph, $i$ is the parent node, $j$ is the child
node, and $j$ is neighboring to $i$.
A graph with
the property that $(i,j)\in\mathcal {E}$ implies $(j,i)\in\mathcal {E}$  is said to be
undirected. A path on $\mathcal
{G}$ from node $i_1$ to node $i_l$ is a sequence of ordered edges of
the form $(i_k, i_{k+1})$, $k=1,\cdots,l-1$. A directed graph
contains a directed spanning tree if there exists a node called the root,
which has no parent,
such that there exists a directed path from this node to every other
node.

The adjacency
matrix $A\in\mathbf{R}^{N\times N}$ of graph $\mathcal {G}$ is defined as
$a_{ii}=0$,
and $a_{ij}>0$ if $(j,i)\in\mathcal {E}$ but $0$ otherwise. The
Laplacian matrix $\mathcal {L}\in\mathbf{R}^{N\times N}$ is
defined as $\mathcal {L}_{ii}=\sum_{j\neq i}a_{ij}$, $\mathcal
{L}_{ij}=-a_{ij}$ for $i\neq j$. Given
a matrix $R = (r_{ij})_{p\times p}$, the graph of $R$ is
the directed graph with $p$ nodes such that there is an edge in the
graph from node $j$ to node $i$ if and only if $r_{ij}\neq 0$
\cite{horn}.

{\small \bf Lemma 1} \cite{ren05}. Zero is an eigenvalue of
$\mathcal {L}$ with $\mathbf{1}$ as the corresponding right
eigenvector and all the nonzero eigenvalues have positive real
parts. Furthermore, zero is a simple eigenvalue of $\mathcal {L}$ if
and only if the graph has a directed spanning tree.

\section{Distributed Adaptive Attitude Synchronization}

This section considers the distributed adaptive attitude
synchronization problem of \dref{rbd} whose communication topology
is represented by a leaderless directed graph $\mathcal {G}$. A
graph is leaderless, if each node in this graph has at least one
parent. Before moving forward, the attitude synchronization problem
is first defined.

{\small \bf Definition 1}. The distributed adaptive attitude synchronization problem
is said to be solved, if the control laws $u_i$, $i=1,2,\cdots,N$, are designed by using only
local
information of neighboring spacecraft such that the attitudes of
\dref{rbd} satisfy $\lim_{t\rightarrow
\infty}\|\sigma_i-\sigma_j\|=0$, $\lim_{t\rightarrow
\infty}\|\dot{\sigma}_i-\dot{\sigma}_j\|=0$, $i,j=1,2,\cdots,N$.


At each time instant, the attitude information of other spacecraft
available to spacecraft $i$ is given by
\begin{equation}\label{atsg}
\sigma_i^d\triangleq \frac{\sum_{j=1}^N a_{ij}\sigma_j}{\sum_{j=1}^N a_{ij}},
\quad i=1,2,\cdots,N,
\end{equation}
where $A=(a_{ij})_{N\times N}$ is the adjacency matrix of the communication graph
$\mathcal {G}$. 

To quantify whether the attitude
synchronization is achieved or not, a synchronization error $e_i(t)$ is defined as follows:
\begin{equation}\label{aerr}
e_i\triangleq \sigma_i-\sigma_i^d, \quad i=1,2,\cdots,N.
\end{equation}
In addition, further define a filtered synchronization error
$s_i(t)$ as
\begin{equation}\label{aerrf}
s_i\triangleq \dot{e}_i+\Lambda_i e_i, \quad i=1,2,\cdots,N,
\end{equation}
where $\Lambda_i\in\mathbf{R}^{3\times 3}$, $i=1,2,\cdots,N,$ are constant
positive-definite matrices.

In light of \dref{rb}, \dref{atsg}, \dref{aerr}, and \dref{aerrf}, it can be
obtained that vector $s_i$ evolves according to the following dynamics:
\begin{equation}\label{aerdy}
H^*_i(\sigma_i)\dot{s}_i+C^*_i(\sigma_i,\dot{\sigma}_i)s_i=
G^{-T}(\sigma_i)u_i-H^*_i(\sigma_i)(\ddot{\sigma}_i^d-\Lambda_i
\dot{e}_i)-
C^*_i(\sigma_i,\dot{\sigma}_i)(\dot{\sigma}_i^d-\Lambda_i e_i),~
i=1,2,\cdots,N.
\end{equation}
By Property 3, the right side of the above equation can be written
into a linear combination of the inertia vector $\theta_i$, which
is defined in \dref{lpm}. To the end, introduce a linear
operator $L(a)$ for vector $a=[a_1,a_2,a_3]^T$ as
$$L(a)=\begin{bmatrix}
a_1 & a_2 & a_3 & 0& 0 &0 \\0 & a_1 & 0 & a_2 & a_3 & 0\\
0 & 0 & a_1 & 0 & a_2 & a_3
\end{bmatrix},$$
and a linear operator $F(x,v)$ for vectors $x=[x_1,x_2,x_2]^T$, $v=[v_1,v_2,v_3]^T$ as
$$F(x,v)=\begin{bmatrix}
0& x_1v_3& -x_1v_2 & x_2v_3 & -x_2v_2+x_3v_3 & -x_3v_2\\
-x_1v_3 & -x_2v_3 & x_1v_1-x_3v_3 & 0 & x_2v_1 & x_3v_1\\
x_1v_2 & -x_1v_1+x_2v_2 & x_3v_2 & -x_2v_1 & -x_3v_1 & 0
\end{bmatrix}.$$
It can be verified that operators $L(s)$ and $F(x,v)$ satisfy
\begin{equation}\label{tran}
J_ia=L(a)\theta_i,\quad S(J_ix)v=F(x,v)\theta_i.
\end{equation}
Therefore, by using \dref{tran}, \dref{aerdy} can be written as
\begin{equation}\label{aerdy2}
H^*_i(\sigma_i)\dot{s}_i+C^*_i(\sigma_i,\dot{\sigma}_i)s_i=
G^{-T}(\sigma_i)u_i-
Y_i(\sigma_i,\dot{\sigma}_i,\dot{\sigma}_i^d-\Lambda_i
e_i,\ddot{\sigma}_i^d-\Lambda_i
 \dot{e}_i)\,\theta_i,\quad
i=1,2,\cdots,N,
\end{equation}
where
$$\begin{aligned}
Y_i(\sigma_i,\dot{\sigma}_i,\dot{\sigma}_i^d-\Lambda_i
e_i,\ddot{\sigma}_i^d-\Lambda_i
 \dot{e}_i)=G^{-T}(\sigma_i)\left(L(G^{-1}(\sigma_i)(\ddot{\sigma}_i^d-\Lambda_i
\dot{e}_i))+L(G^{-1}(\sigma_i)(\dot{\sigma}_i^d-\Lambda_i e_i))\right.\\
+\left.F(G^{-1}(\sigma_i)\dot{\sigma}_i,G^{-1}(\sigma_i)(\dot{\sigma}_i^d-\Lambda_i e_i))\right)
\end{aligned}$$

Since the inertia parameter $\theta_i$ is unknown, its estimate $\hat{\theta}_i$
is used instead to construct the controller
$u_i$ to spacecraft $i$ as follows:
\begin{equation}\label{con}
u_i=G^{T}(\sigma_i)\left(Y_i(\sigma_i,\dot{\sigma}_i,\dot{\sigma}_i^d-\Lambda_i
e_i,\ddot{\sigma}_i^d- \Lambda_i
\dot{e}_i)\,\hat{\theta}_i-K_is_i\right), \quad i=1,2,\cdots,N,
\end{equation}
where $K_i\in\mathbf{R}^{3\times 3}$, $i=1,2,\cdots,N,$ are positive
definite. The parameter estimate vector $\hat{\theta}_i$ is
generated by the following adaptive updating law:
\begin{equation}\label{upd}
\dot{\hat{\theta}}_i=-\Gamma_i
Y_i^T(\sigma_i,\dot{\sigma}_i,\dot{\sigma}_i^d-\Lambda_i
e_i,\ddot{\sigma}_i^d-\Lambda_i \dot{e}_i) s_i,\quad i=1,2,\cdots,N,
\end{equation}
where $\Gamma_i\in\mathbf{R}^{6\times 6}$, $i=1,2,\cdots,N,$ are positive-definite
diagonal matrices.

Define the estimation errors
$\tilde{\theta}_i=\theta_i-\hat{\theta}_i$, $i=1,2,\cdots,N$. Then,
substituting \dref{con}, \dref{upd} into \dref{aerdy2} gives
\begin{equation}\label{ady}
H^*_i(\sigma_i)\dot{s}_i+C^*_i(\sigma_i,\dot{\sigma}_i)s_i= -
Y_i(\sigma_i,\dot{\sigma}_i,\dot{\sigma}_i^d-\Lambda_i
e_i,\ddot{\sigma}_i^d-\Lambda_i \dot{e}_i)\,\tilde{\theta}_i
-K_is_i,\quad i=1,2,\cdots,N.
\end{equation}

{\small \bf Theorem 1}. If the leaderless communication graph
$\mathcal {G}$ has a directed spanning tree, then distributed controllers
\dref{con} and adaptive updating laws
\dref{upd} solves the attitude synchronization problem for \dref{rbd}.

{\small \bf Proof}. Let $e(t)=[e_1^T,\cdots,e_N^T]^T$ and
$\sigma(t)=[\sigma_1^T,\cdots,\sigma_N^T]^T$. Then, \dref{aerr} can
be written as
\begin{equation}\label{aerrs1}
e=M\mathcal {L}\sigma,
\end{equation}
where $\mathcal {L}$ is the
Laplacian matrix associated with graph $\mathcal {G}$, 
and $M=\rm{diag}(\sum_{j=1}^N a_{1j},\cdots,\sum_{j=1}^N a_{Nj})$.
Since graph $\mathcal {G}$ is leaderless and has a directed spanning tree,
one obtains 1) there exists at least one nonzero entry for each row
of the adjacency matrix $A$, thus matrix $M$ is positive definite;
2) the Laplacian matrix $\mathcal {L}$ has a simple eigenvalue $0$
with $\mathbf{1}$ as the corresponding eigenvector, and the other
eigenvalues have positive real parts. Then, it follows from
\dref{aerrs1} that $e=0$ if and only if
$\sigma_1=\sigma_2=\cdots=\sigma_N$. That is, the attitude
synchronization problem is solved if and only if $e(t)\rightarrow
0$, $\dot{e}(t)\rightarrow 0$, as $t\rightarrow\infty$.

Consider the following Lyapunov function candidate
\begin{equation}\label{lya}
V(t)=\frac{1}{2}\left(\sum_{i=1}^N
s_i^TH_i^*(\sigma_i)s_i+\sum_{i=1}^N
\tilde{\theta}_i^T\Gamma_i^{-1}\tilde{\theta}_i\right).
\end{equation}
By \dref{ady}, \dref{upd}, and Property 2, the time derivative of
$V(t)$ is
\begin{equation}\label{lyad}
\begin{aligned}
\dot{V}(t) &=\sum_{i=1}^N
s_i^T\left(\dot{H}_i^*(\sigma_i)s_i+H_i^*(\sigma_i)\dot{s}_i\right)+\sum_{i=1}^N
\dot{\tilde{\theta}}_i^T\Gamma_i^{-1}\tilde{\theta}_i\\
&=\sum_{i=1}^N
s_i^T\left(\dot{H}_i^*(\sigma_i)s_i-C_i^*(\sigma_i,\dot{\sigma}_i)-Y_i(\sigma_i,\dot{\sigma}_i,\dot{\sigma}_i^d-\Lambda_i
e_i,\ddot{\sigma}_i^d-\Lambda_i \dot{e}_i)\tilde{\theta}_i
-K_is_i\right)\\&\quad-\sum_{i=1}^N
\dot{\hat{\theta}}_i^T\Gamma_i^{-1}\tilde{\theta}_i\\
&=-\sum_{i=1}^N
\left(s_i^TY_i(\sigma_i,\dot{\sigma}_i,\dot{\sigma}_i^d-\Lambda_i
e_i,\ddot{\sigma}_i^d-\Lambda_i \dot{e}_i)\tilde{\theta}_i
+\dot{\hat{\theta}}_i^T\Gamma_i^{-1}\tilde{\theta}_i+s_i^TK_is_i\right)\\
&=-\sum_{i=1}^N  s_i^TK_is_i\leq 0.
\end{aligned}
\end{equation}

Let $S=\{(\sigma_i,s_i,\tilde{\theta}_i)|\dot{V}=0\}$. Note that $\dot{V}=0$ implies that $s_i=0$,
$i=1,2,\cdots,N$.
By LaSalle's invariance principle \cite{slotine}, it follows from that
$s_i\rightarrow 0$, $i=1,2,\cdots,N$, as $t\rightarrow \infty$,
which by \dref{aerrf} in turn shows that $e(t)\rightarrow 0$,
$\dot{e}(t)\rightarrow 0$, as $t\rightarrow\infty$, i.e,
attitude synchronization is achieved. \hfill $\blacksquare$

{\small \bf Remark 1}.
It should be noted that the controller \dref{con} to spacecraft $i$ depends only on
its own attitude vectors $\sigma_i$, $\dot{\sigma}_i$, $\ddot{\sigma}_i$, and the attitudes
of its neighboring spacecraft, therefore is distributed. Theorem 1 gives a sufficient condition for
achieving attitude synchronization. However, it is generally quite hard to expressly derive
the final synchronized attitude value, which
depends on the initial attitudes of the $N$ spacecraft,
matrices $\Lambda_i$, $K_i$, $i=1,2,\cdots,N$, and the
communication topology $\mathcal {G}$.

\section{Distributed Adaptive Attitude Tracking}

%

Different from the leaderless communication graph discussed as in
the above section, the spacecraft' attitudes may be desired to
follow a given time-varying reference attitude $\sigma^r$ in certain
circumstance. It is supposed that $\sigma^r$ is available to only a
subgroup of the spacecraft, otherwise cooperation between
neighboring spacecraft by exchanging attitude information become
not so necessary. Assume that $\sigma^r$, $\dot{\sigma}^r$, and
$\dot{\sigma}^r$ are all bounded. For this case, the attitude
synchronization problem is called attitude tracking problem in
\cite{chung,ren10}, which is formulated as follows.

{\small \bf Definition 2}. The distributed adaptive attitude
tracking problem is said to be solved, if the local control laws
$u_i$, $i=1,2,\cdots,N$, are designed for \dref{rbd} such that
$\lim_{t\rightarrow \infty}\|\sigma_i-\sigma^r\|=0$,
$\lim_{t\rightarrow \infty}\|\dot{\sigma}_i-\dot{\sigma}^r\|=0$,
$i=1,2,\cdots,N$.

Take the
reference attitude $\sigma^r$ as the attitude of a virtual leader, labeled as spacecraft $N+1$.
Since the virtual leader does not obtain
any information from the $N$ spacecraft, the communication
topology among these $N+1$ spacecraft (the $N$ spacecraft and the virtual leader)
is in the leader-follower form.

Assume that the communication topology among the $N$ spacecraft is
still denoted by $\mathcal {G}$. The attitude information of other
spacecraft available to spacecraft $i$ is given by
\begin{equation}\label{atsgf}
\sigma_{fi}^{d}\triangleq \frac{\sum_{j=1}^N
a_{ij}\sigma_j+a_{i(N+1)}\sigma^r}{\sum_{j=1}^N a_{ij}+a_{i(N+1)}},
\quad i=1,2,\cdots,N,
\end{equation}
where $A=(a_{ij})_{N\times N}$ is the adjacent matrix of $\mathcal {G}$,
$a_{i(N+1)}>0$, $i=1,\cdots,N$, if spacecraft $i$ has access to the virtual leader
and $a_{i(N+1)}=0$ otherwise.
Similar to the above section, a synchronization error $e_{fi}(t)$ is defined as follows:
\begin{equation}\label{aerrfl}
\dot{e}_{fi}=\sigma_{fi}-\sigma_{fi}^{d}, \quad i=1,2,\cdots,N.
\end{equation}
Correspondingly, the filtered synchronization errors $s_{fi}$ is
defined as
\begin{equation}\label{aerrf2}
s_{fi}\triangleq \dot{e}_{fi}+\widehat{\Lambda}_i e_{fi}, \quad
i=1,2,\cdots,N,
\end{equation}
where matrices $\widehat{\Lambda}_i\in\mathbf{R}^{3\times 3}$,
$i=1,2,\cdots,N,$ are positive definite. By \dref{rb}, \dref{atsgf},
\dref{aerrfl}, and Property 3, vector $s_{fi}$ satisfies the
following dynamics:
\begin{equation}\label{aerdyf}
H^*_i(\sigma_i)\dot{s}_{fi}+C^*_i(\sigma_i,\dot{\sigma}_i)s_{fi}=
G^{-T}(\sigma_i)u_i-
Y_i(\sigma_i,\dot{\sigma}_i,\dot{\sigma}_{fi}^{d}-\widehat{\Lambda}_i
e_{fi},\ddot{\sigma}_{fi}^{d}-\widehat{\Lambda}_i
 \dot{e}_{fi})\,\theta_i,\quad
i=1,2,\cdots,N.
\end{equation}

The distributed controllers $u_i$, $i=1,2,\cdots,N,$ to the $N$
spacecraft are proposed as
\begin{equation}\label{conlf}
\begin{aligned}
u_i
=G^{T}\left(Y_i(\sigma_i,\dot{\sigma}_i,\dot{\sigma}_{fi}^{d}-\widehat{\Lambda}_i
e_{fi},\ddot{\sigma}_{fi}^{d}- \widehat{\Lambda}_{i}
\dot{e}_{fi})\,\hat{\theta}_{fi}-\widehat{K}_is_{fi}\right), \quad
i=1,2,\cdots,N,
\end{aligned}
\end{equation}
where $\widehat{K}_i\in\mathbf{R}^{3\times 3}$, $i=1,2,\cdots,N,$
are positive definite and $\hat{\theta}_{fi}$ is the estimate of $\theta_i$,
generated by the following adaptive updating law:
\begin{equation}\label{updf}
\dot{\hat{\theta}}_{fi}=-\widehat{\Gamma}_i
Y_i^T(\sigma_i,\dot{\sigma}_i,\dot{\sigma}_{fi}^{d}-\widehat{\Lambda}_i
e_{fi},\ddot{\sigma}_{fi}^{d}- \widehat{\Lambda}_i
\dot{e}_{fi})s_{fi},\quad i=1,2,\cdots,N,
\end{equation}
where $\widehat{\Gamma}_i\in\mathbf{R}^{6\times 6}$,
$i=1,2,\cdots,N,$ are positive-definite diagonal matrices. 

{\small \bf Theorem 2}. Denote by $\widehat{\mathcal {G}}_{N+1}$ the
directed graph of matrix $A_{N+1}=\begin{bmatrix} A & b\\ 0 &
0\end{bmatrix}$, where $b=[a_{1(N+1)},\cdots,a_{N(N+1)}]^T$. If
graph $\widehat{\mathcal {G}}_{N+1}$ has a spanning tree with node
$N+1$ as the root, then distributed controllers \dref{conlf} and
adaptive updating laws \dref{updf} solve the attitude tracking
problem for the $N$
spacecraft in \dref{rbd}. 

{\small \bf Proof}. Let $\sigma_{N+1}\triangleq\sigma^r$,
$\hat{\sigma}(t)=[\sigma_1^T,\cdots,\sigma_N^T,\sigma_{N+1}^T]^T$,
and $e_f(t)=[e_{f1}^T,\cdots,e_{fN}^T,0]^T$. Then, \dref{aerrfl} can
be written as
\begin{equation}\label{aerrsf1}
e_f=\widehat{M}\widehat{\mathcal {L}}\hat{\sigma},
\end{equation}
where $\widehat{M}=\rm{diag}(\sum_{j=1}^N a_{1j},\cdots,\sum_{j=1}^N
a_{Nj},1)$, and $\widehat{\mathcal {L}}=(\widehat{\mathcal {L}}_{ij})_{(N+1)\times
(N+1)}$ is the Laplacian matrix of graph $\widehat{\mathcal {G}}$,
defined as $\widehat{\mathcal {L}}_{ii}=\sum_{j=1}^{N+1}a_{ij}$,
$\widehat{\mathcal {L}}_{ij}=-a_{ij}\,(j\neq i)$,
$\forall\,i\in\{1,2,\cdots,N\},$ and $\widehat{\mathcal
{L}}_{(N+1)j}=0$, $\forall\,j\in\{1,2,\cdots,N+1\}$. Under the
assumption of the theorem, matrix $\widehat{M}$ is positive definite
and $0$ is a simple eigenvalue of matrix $\widehat{\mathcal {L}}$
with $\mathbf{1}$ as the corresponding eigenvector, implying that
$e_f=0$ if and only if $\sigma_1=\sigma_2=\cdots=\sigma_N=\sigma^r$.
That is, the distributed attitude tracking problem is solved if and
only if $e_f(t)\rightarrow 0$, $\dot{e}_f(t)\rightarrow 0$, as
$t\rightarrow\infty$.

Consider the following Lyapunov function candidate
\begin{equation}\label{lyaf}
\widehat{V}(t)=\frac{1}{2}\left(\sum_{i=1}^N
s_{fi}^TH_i^*(\sigma_i)s_{fi}+\sum_{i=1}^N
(\theta_i-\hat{\theta}_{fi})^T\widehat{\Gamma}_i^{-1}(\theta_i-\hat{\theta}_{fi})\right).
\end{equation}
The time derivative of $\widehat{V}(t)$ can be obtained as
\begin{equation}\label{lyadf}
\begin{aligned}
\dot{\widehat{V}}(t) &=\sum_{i=1}^N
s_{fi}^T\left(\dot{H}_i^*(\sigma_i)s_{fi}+H_i^*(\sigma_i)\dot{s}_{fi}\right)-\sum_{i=1}^N
\dot{\hat{\theta}}_{fi}^T\widehat{\Gamma}_i^{-1}(\theta_i-\hat{\theta}_{fi})\\
&=-\sum_{i=1}^N  s_{fi}^T\widehat{K}_is_{fi}.
\end{aligned}
\end{equation}

Since $\widehat{V}(t)\geq 0$, $\dot{\widehat{V}}(t)\leq 0$,
$\widehat{V}(t)$ remains bounded, which by \dref{lyaf} implies that
$s_{fi}$, $\tilde{\theta}_{fi}$, $i=1,2,\cdots,N$, are bounded. By
using standard signal chasing arguments, it is easy to show that
$\dot{s}_{fi}$, $i=1,2,\cdots,N$, are bounded. Therefore,
$\ddot{\widehat{V}}(t)= -2\sum_{i=1}^N
\dot{s}_{fi}^T\widehat{K}_is_{fi}$ is bounded. In light
of Barbalat's lemma \cite{slotine}, $\dot{\widehat{V}}(t)\rightarrow
0$, as $t\rightarrow\infty$. Thus, $s_i\rightarrow 0$,
$i=1,2,\cdots,N$, as $t\rightarrow \infty$, which by \dref{aerrfl}
shows that $e_f(t)\rightarrow 0$, $\dot{e}_f(t)\rightarrow 0$, as
$t\rightarrow\infty$, i.e, the distributed attitude tracking problem
is solved. \hfill $\blacksquare$

{\small \bf Remark 2.} Theorems 1 and 2 present sufficient conditions for the adaptive
attitude synchronization of multiple spacecraft having general
directed communications in the presence of unknown inertia matrices,
for both the cases with and without a reference attitude. By
contrast, the results given in \cite{chung} are applicable only to a
bidirectional ring communication topology, and the result in
\cite{cheng} requires the communication graph to be undirected.
Theorems 1 and 2 generalize Theorem 4.1 in \cite{ren10} to the case
where the attitude dynamics \dref{rbd} are uncertain and the
communication topology is either leaderless or leader-follower. It
should be noted that different from Theorem 1, Barbalat's lemma is
utilized to derive Theorem 2, due to the fact that the time-varying
reference attitude $\sigma^r$ may render the closed-loop system
nonautonomous.

\section{Numerical Examples}

In this section, the effectiveness of the proposed control laws is
illustrated through numerical simulations.

\begin{figure}[htbp]
\centering
\includegraphics[width=0.3\linewidth]{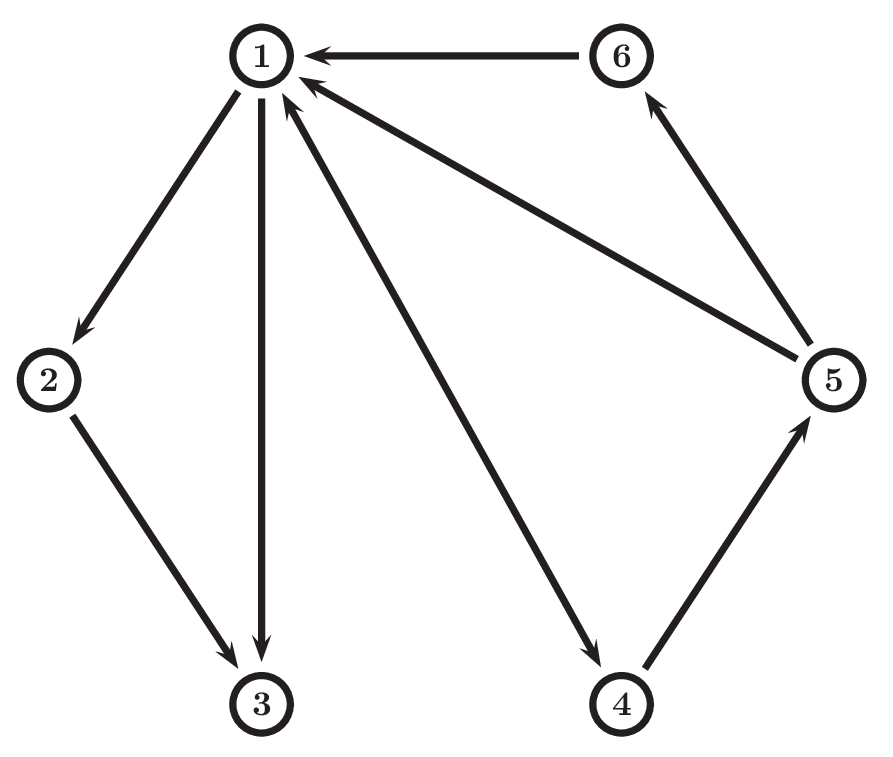}
\caption{The communication topology. }
\end{figure}

Consider a group of six spacecraft, whose inertia matrices are shown in Table 1 \cite{ren071}.
The initial attitude states $\sigma_i(0)$, $\omega_i(0)$,
$i=1,2,\cdots,6$, are chosen randomly.
The communication topology is given by
Fig. 1, so the corresponding adjacency matrix is
$$A=\begin{bmatrix} 0 & 0 & 0 & 1 & 1 & 1\\ 1 & 0 & 0 &0 &0 &0\\
1 & 1 & 0 & 0 & 0& 0\\ 1 & 0 & 0 & 0 & 0 & 0\\ 0 & 0 & 0 & 1 & 0
& 0\\ 0 & 0 & 0 & 0 & 1 & 0\end{bmatrix}.$$
For \dref{con} and \dref{conlf}, take matrices $\Lambda_i=\widehat{\Lambda}_i=I_3$,
$K_i=\widehat{K}_i=3I_3$,
$\Gamma_i=\widehat{\Gamma}_i=3I_6$, $i=1,2,\cdots,6$.
The parameter estimates $\hat{\theta}_i$ and $\hat{\theta}_{fi}$
are initialized to be zero, i.e., $\hat{\theta}_i(0)=0$, $\hat{\theta}_{fi}(0)=0$,
$i=1,2,\cdots,6$, in \dref{upd}, \dref{updf}. For simplicity, let the reference attitude
$\sigma^r=[0.1, 0.3, 0.5]^T$. Suppose
that $\sigma^r$ is available only to spacecraft 1. In this case, $a_{17}=1$,
$a_{i7}=0$, $i=2,\cdots,6$.

\begin{table}[htbp]
\centering
\begin{tabular}{ll}
\hline
$J_1\qquad$ & $[1, 0.1, 0.1; 0.1, 0.1, 0.1; 0.1, 0.1, 0.9]$  kgm$^2$\\
$J_2\qquad$ & $[1.5, 0.2, 0.3; 0.2, 0.9, 0.4; 0.3, 0.4, 2.0]$  kgm$^2$\\
$J_3\qquad$ & $[0.8, 0.1, 0.2; 0.1, 0.7, 0.3; 0.2, 0.3, 1.1]$  kgm$^2$\\
$J_4\qquad$ & $[1.2, 0.3, 0.7; 0.3, 0.9, 0.2; 0.7, 0.2, 1.4]$  kgm$^2$\\
$J_5\qquad$ & $[0.9, 0.15, 0.3; 0.15, 1.2, 0.4; 0.3, 0.4, 1.2]$ kgm$^2$\\
$J_6\qquad$  & $[1.1, 0.35, 0.45; 0.35, 1.0, 0.5; 0.45, 0.5, 1.3]$ kgm$^2$\\
\hline
\end{tabular}
\caption{Spacecraft specifications.}
\end{table}

Figs. 2(a), 2(b), and 2(c) depict, respectively, the attitudes, angular velocities, and control torques of
the six spacecraft with \dref{con} and \dref{upd}, from which it can be
observed that attitude synchronization is indeed achieved.
The parameter estimates $\hat{\theta}_i$, $i=1,2,\cdots,6$, are shown in Fig. 3.
Figs. 4(a), 4(b), 4(c), and 5 depict, respectively, the attitudes, angular velocities, control torques, and
the parameter estimates $\hat{\theta}_{fi}$, $i=1,2,\cdots,6$, of
the six spacecraft with \dref{conlf} and \dref{updf}.

\begin{figure}[htbp]
\subfigure[Attitudes]{ \begin{minipage}[b]{0.5\linewidth} \centering
\includegraphics[width=\linewidth]{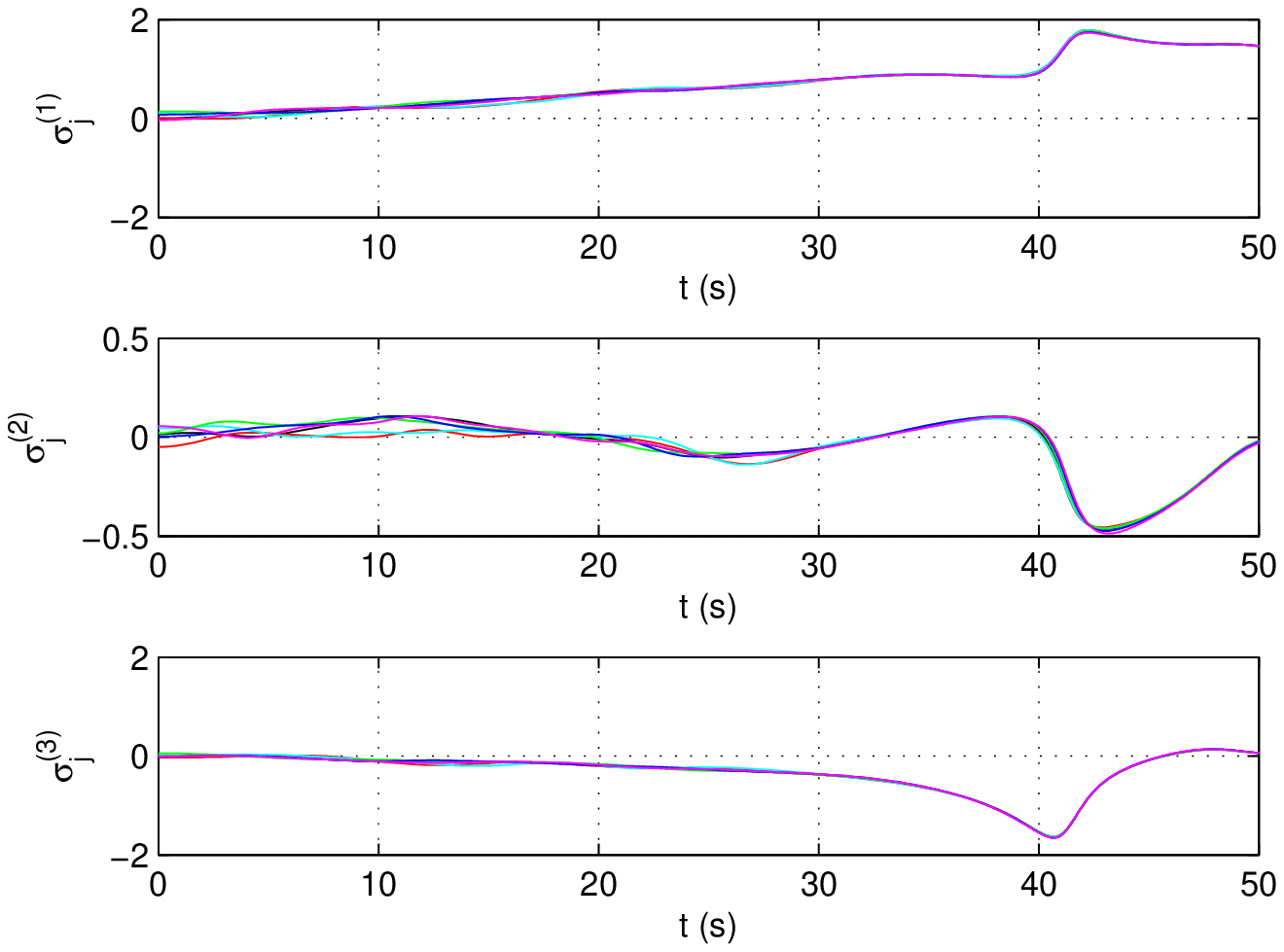}
\end{minipage}}%
\subfigure[Angular velocities]{\begin{minipage}[b]{0.5\linewidth} \centering
\includegraphics[width=\linewidth]{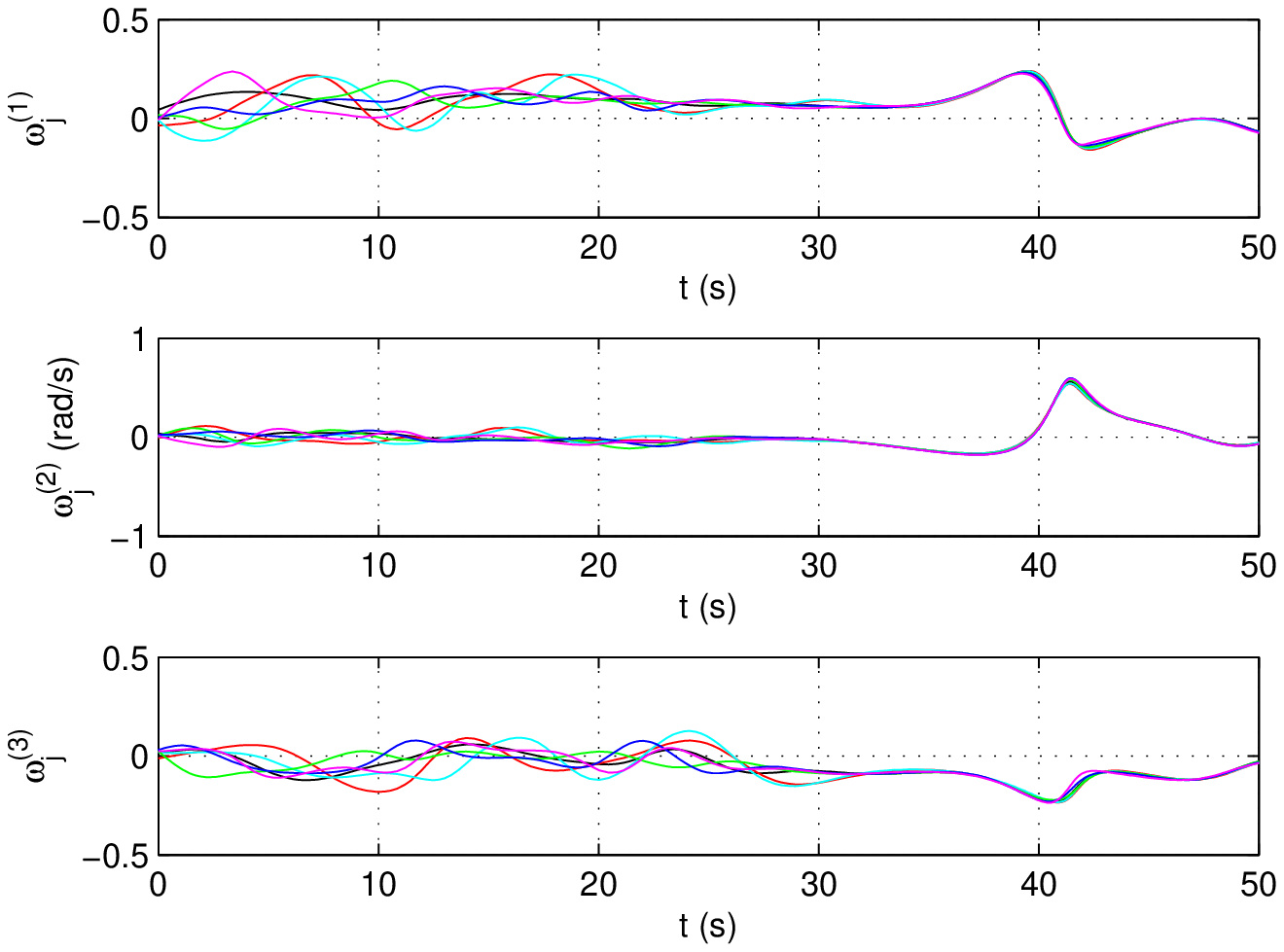}
\end{minipage}}\\
\subfigure[Control torques]{
\includegraphics[width=0.5\linewidth]{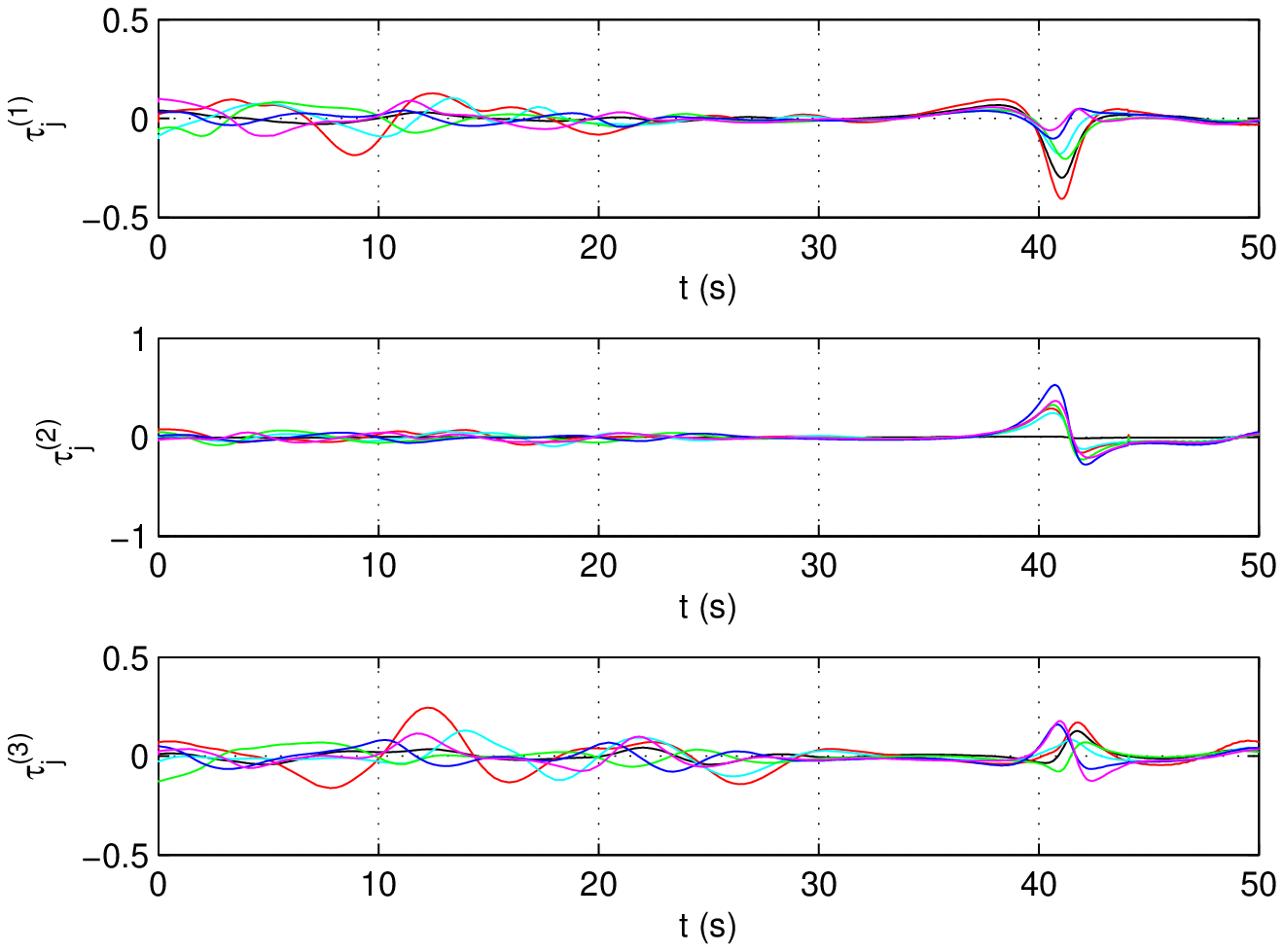}}
\caption{Attitudes, angular velocities, and
control torques of \dref{rbd} with controller \dref{con} and adaptive law \dref{upd}.}
\end{figure}

\begin{figure}[htbp]
\subfigure[$\hat{\theta}_i$, $i=1,2,3$]{ \begin{minipage}[b]{0.5\linewidth} \centering
\includegraphics[width=\linewidth]{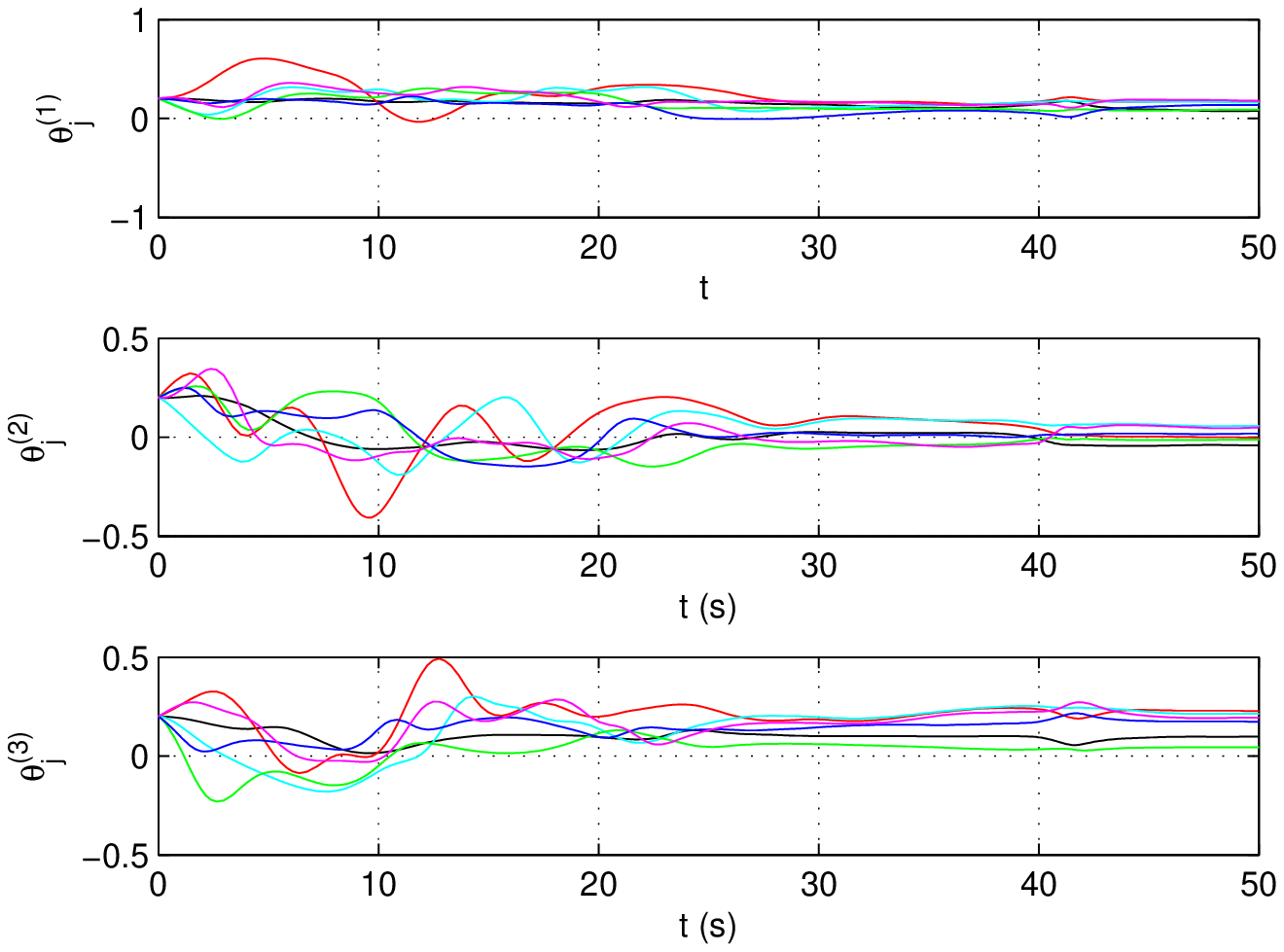}
\end{minipage}}%
\subfigure[$\hat{\theta}_i$, $i=4,5,6$]{\begin{minipage}[b]{0.5\linewidth} \centering
\includegraphics[width=\linewidth]{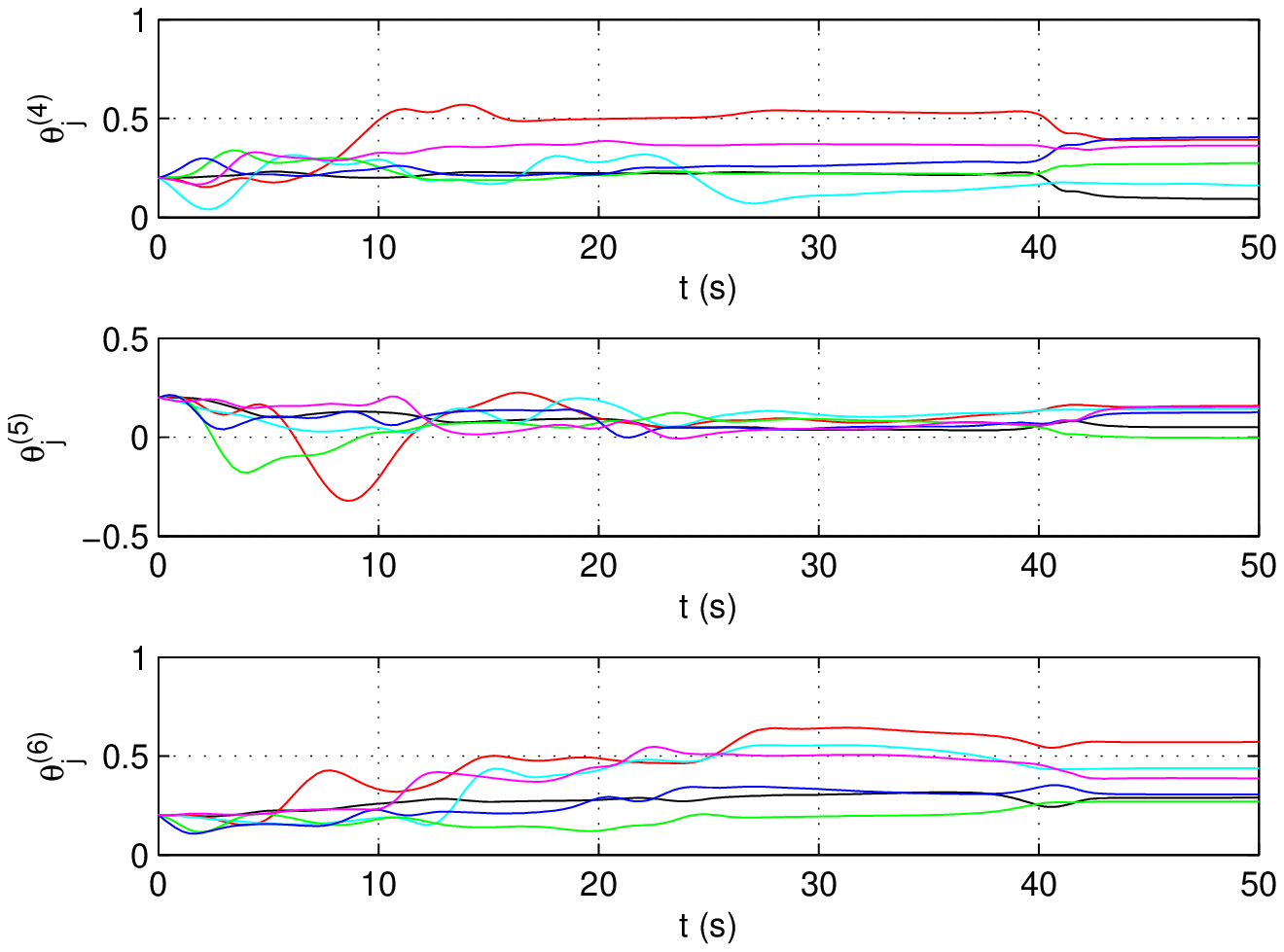}
\end{minipage}}
\caption{Parameter estimates of \dref{upd}.}
\end{figure}

\begin{figure}[htbp]
\subfigure[Attitudes]{ \begin{minipage}[b]{0.5\linewidth} \centering
\includegraphics[width=\linewidth]{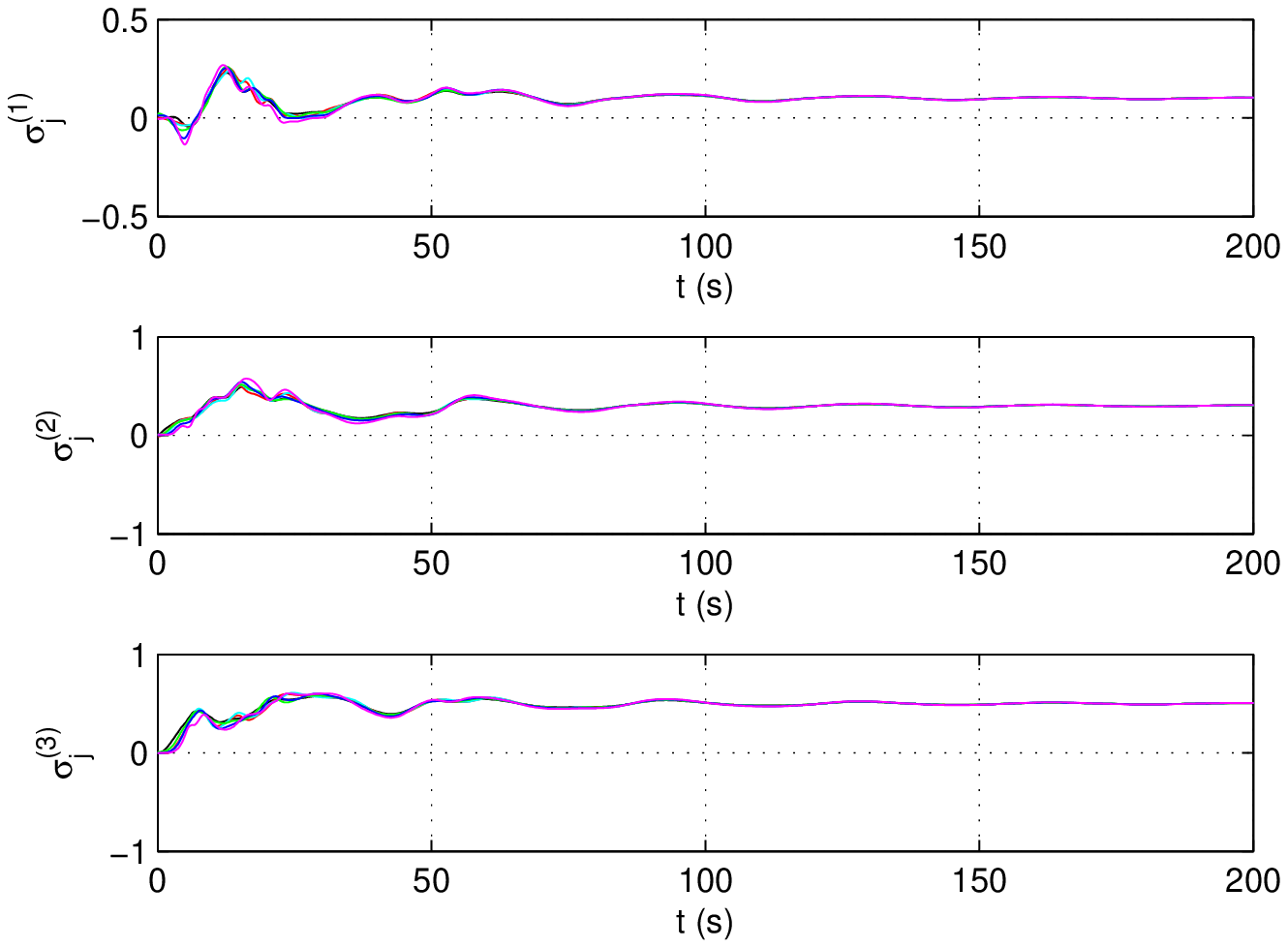}
\end{minipage}}%
\subfigure[Angular velocities]{\begin{minipage}[b]{0.5\linewidth} \centering
\includegraphics[width=\linewidth]{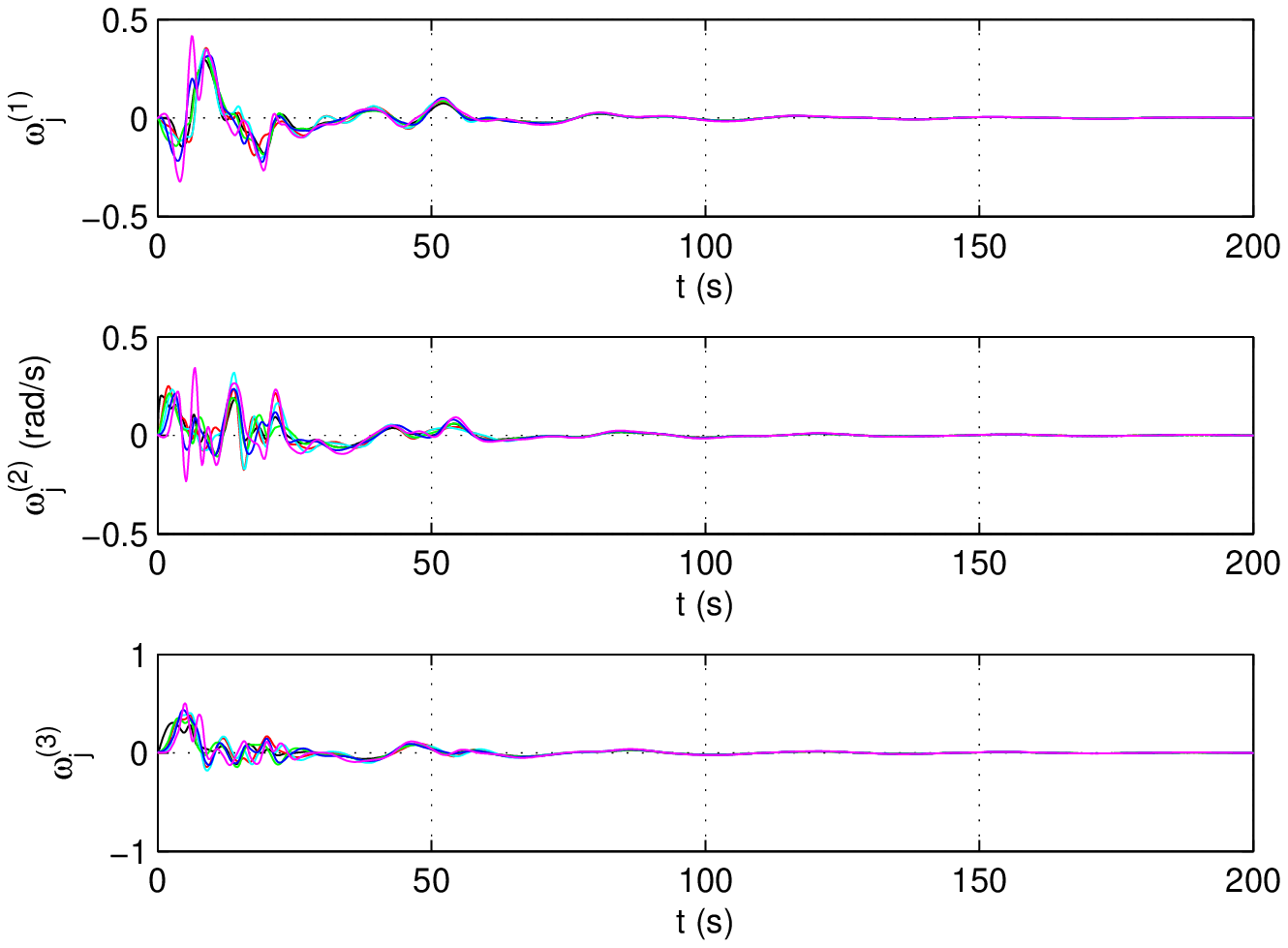}
\end{minipage}}\\
\subfigure[Control torques]{
\includegraphics[width=0.5\linewidth]{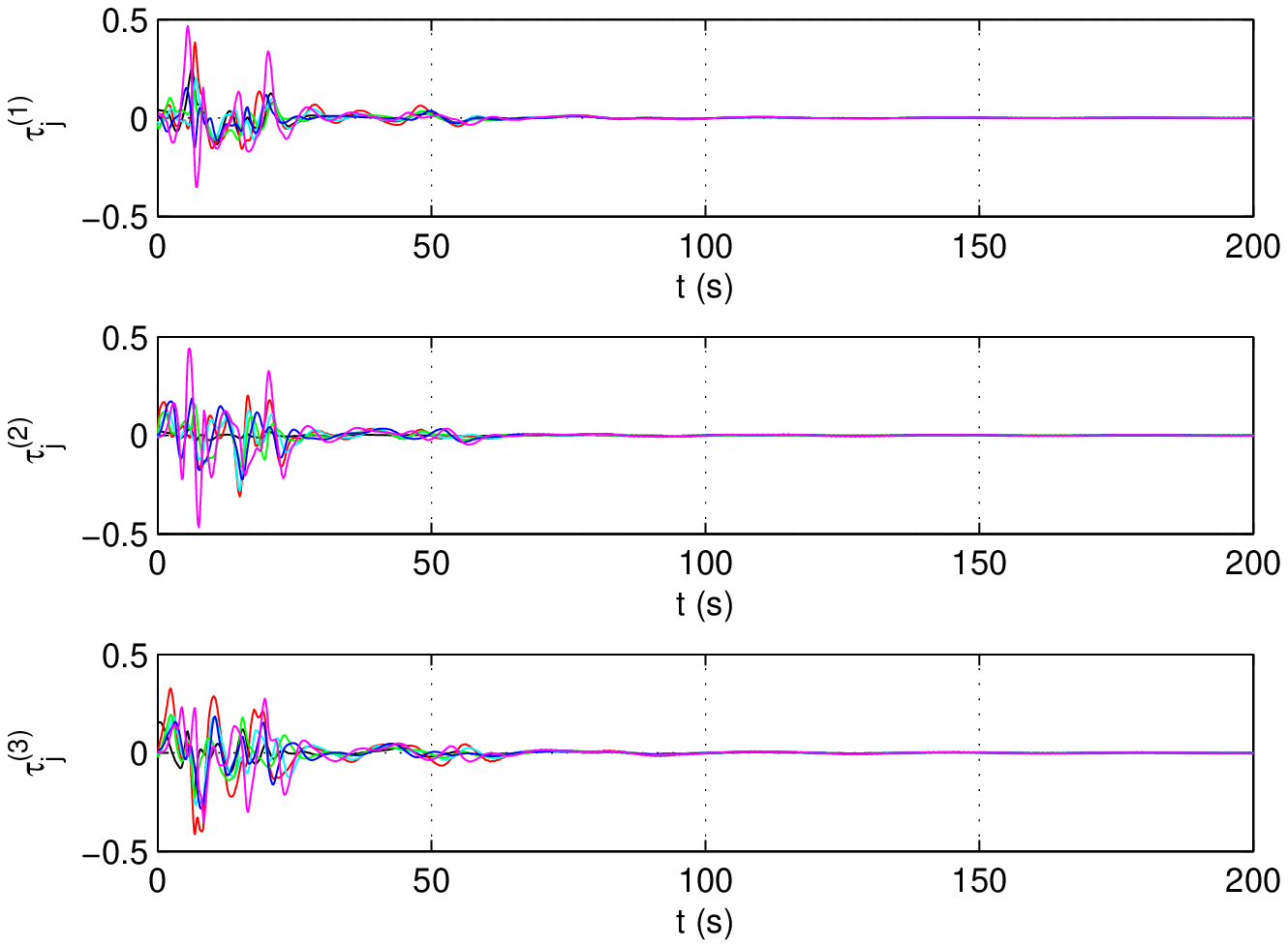}}
\caption{Attitudes, angular velocity, and
control torques of \dref{rbd} with controller \dref{conlf} and adaptive law \dref{updf}.}
\end{figure}

\begin{figure}[htbp]
\subfigure[$\hat{\theta}_{fi}$, $i=1,2,3$]{ \begin{minipage}[b]{0.5\linewidth} \centering
\includegraphics[width=\linewidth]{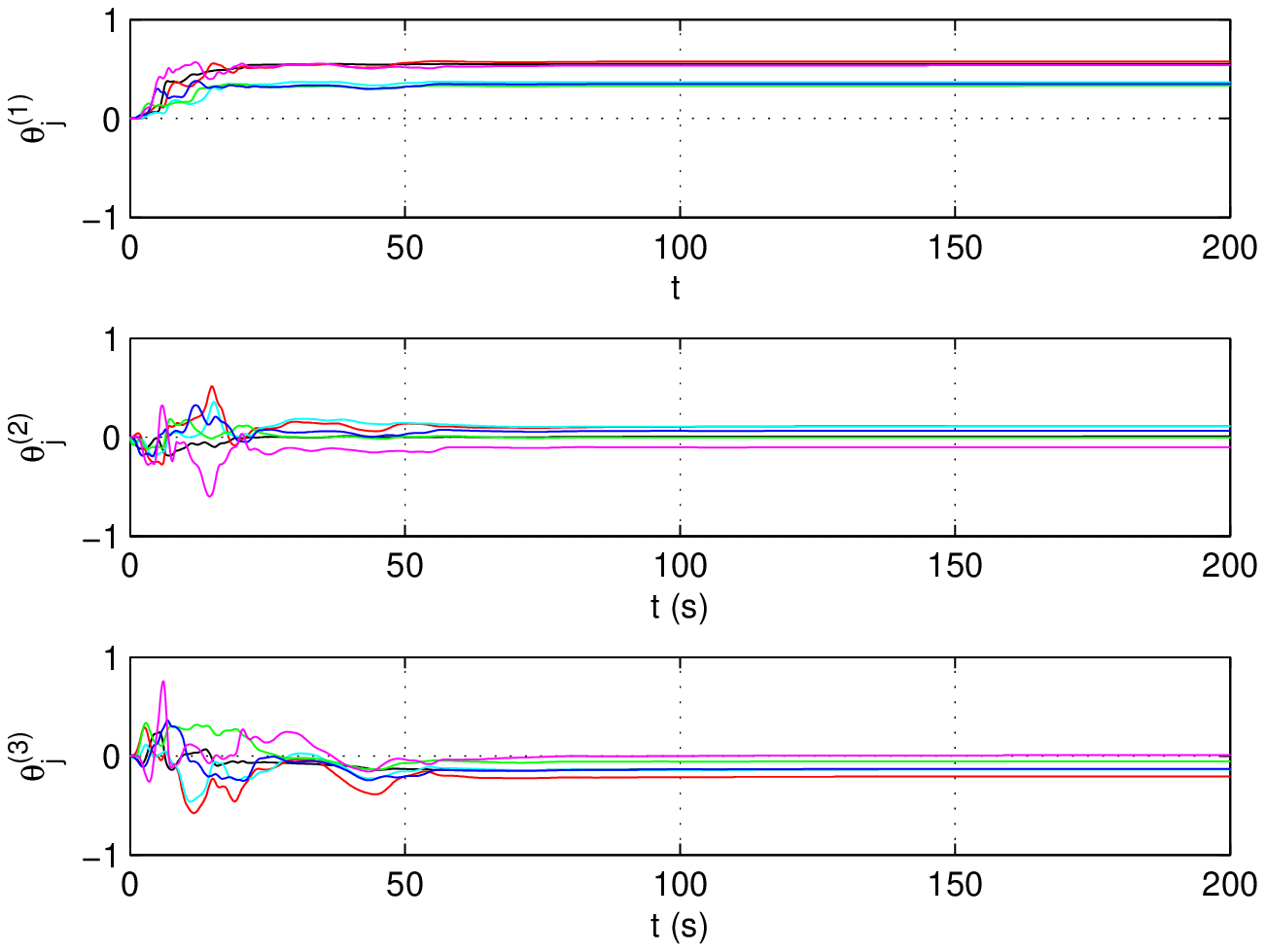}
\end{minipage}}%
\subfigure[$\hat{\theta}_{fi}$, $i=4,5,6$]{\begin{minipage}[b]{0.5\linewidth} \centering
\includegraphics[width=\linewidth]{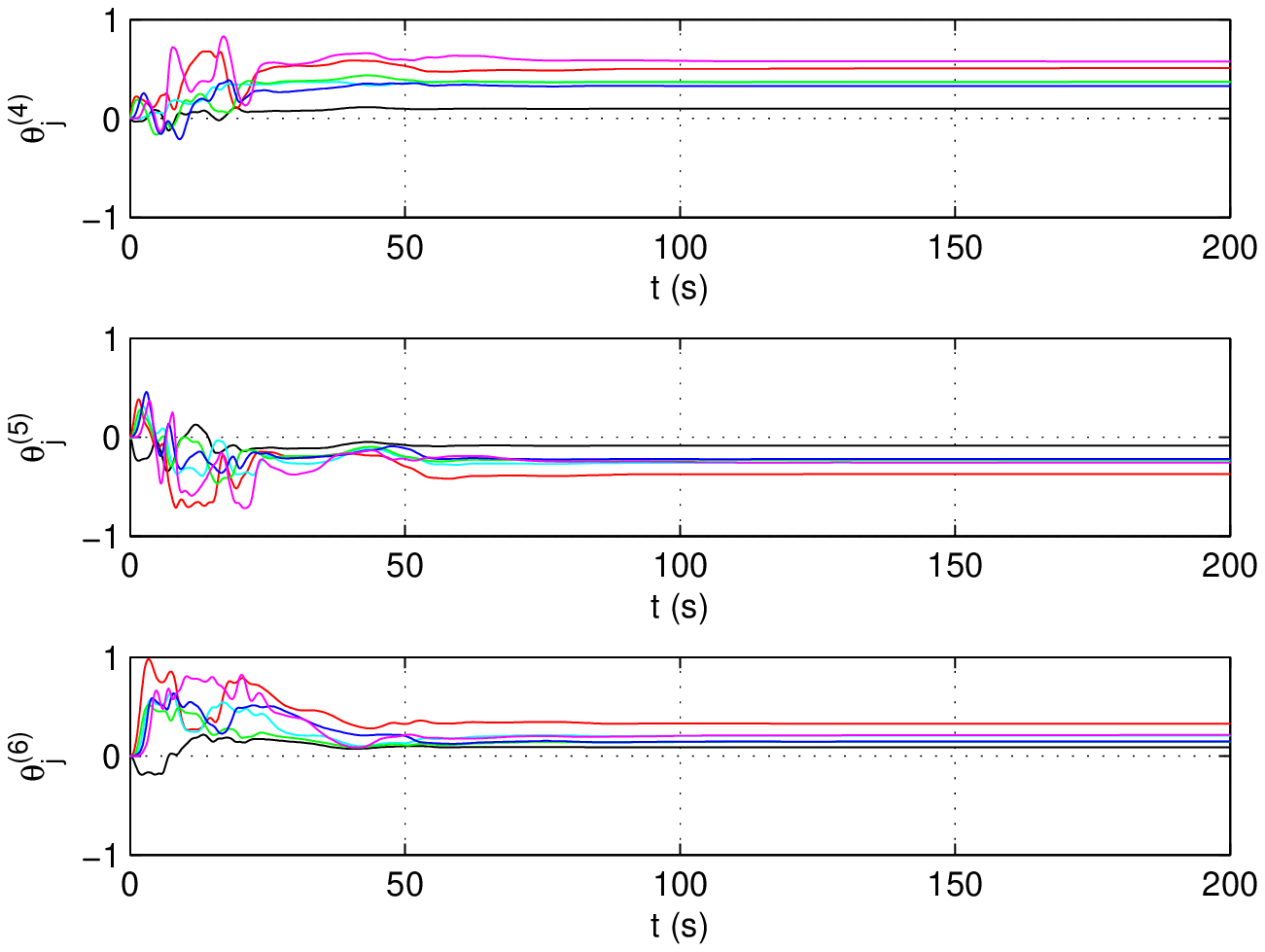}
\end{minipage}}
\caption{Parameter estimates of \dref{updf}. }
\end{figure}

\section{Conclusions}

This paper has addressed the distributed adaptive attitude
synchronization problem of a group of spacecraft with unknown
inertia matrices. Two distributed adaptive controllers have been
proposed for the cases with and without a virtual leader to which a
time-varying reference attitude is assigned. The first controller
achieves attitude synchronization for a group of spacecraft with a
leaderless communication topology having a directed spanning tree.
The second controller guarantees that all spacecraft track the
reference attitude if the virtual leader has a directed path to all
other spacecraft. This paper has extended some existing results in
the literature. An interesting topic for future research is the
distributed adaptive attitude synchronization of multiple
spacecraft without velocity measurements.

\end{document}